\documentclass[twocolumn,floatfix,showpacs,preprintnumbers,amsmath,amssymb,aps,prl]{revtex4}
\usepackage{amssymb,amsmath,graphics,epsfig,amssymb}
\usepackage{color}
\usepackage{color}
\usepackage{wrapfig}

\begin{document}

\title{ A magneto-optical trap created by the 2nd-order external cavity diode lasers}
\author{Jianing Han \footnote{jhan@southalabama.edu}, 
Lindsay Hutcherson, Gayatri Deshmukh, Morgan Umstead, Andy Hu, Young Lee,  Zhanguo Bai, and Juliet Mitchell} 
\affiliation{Physics Department, University of South Alabama, Mobile, AL 36688, USA}

\date{\today}
\begin{abstract}
\label{abst}  
In this article, we report on a magneto-optical trap (MOT) created by the 2nd-order external cavity diode lasers (ECDLs). The lasers were characterized. We have observed the non-continuous changes of the wavelength as a function of the laser diode current. This study is beneficial for achieving tunable atom-atom interactions, quantum tunneling, precision measurement, ultracold plasma, as well as quantum computing.
\end{abstract}

\maketitle
\section{Introduction}
Laser cooling and trapping have been intensively studied in the past few decades (see Ref. \cite{Wineland,Neuhauser, William, Chu, Ketterle, Cornell, Hulet, Jin, Weiss, Anderson, Wenhui} and references therein). A narrowband laser MOT offers greater tunability of the MOT's temperature. In this article, we report on a MOT created by the 2nd-order diffraction external-cavity diode lasers (ECDLs). The second-order Littrow configuration has been studied in the past \cite{Britzger}, and we apply such configurations to laser cooling applications.

Cold atoms offer platforms for a lot of exciting physics \cite{Adrian, Johnston, smith, Han} due to their low kinetic energies. The temperature of the atoms in a MOT is about 300 $\mu$K, and the temperature can be much lower by turning off the quadrupole magnetic field to get optical molasses \cite{Paul}. Table \ref{Tcompare} shows the properties of ultracold atoms compared with the room temperature atoms. First of all, the kinetic energy of the room temperature atom is about 10$^6$ times greater than the kinetic energy of the ultracold atoms. It is shown that the kinetic energy of the room temperature atoms is about 10 THz, while the temperature of ultracold atoms is about 10 MHz. Here we write the energy, $E$, in terms of the frequency, since
\begin{equation}
\begin{split}
E=h\nu,
\end{split}
\label{E_photon}
\end{equation}
where $\nu$ is frequency. The speed of room-temperature atoms is 10$^3$ times greater than that of the ultracold atoms. The last comparison between the room temperature atoms and the ultracold atoms is the distance traveled in one microsecond. The ultracold atoms move 0.3 $\mu$m per microsecond. The slow speed allows us to detect the atoms with less uncertainties.
\begin{table}
\centering
\caption{The comparison between room temperature atoms and the cold atoms in a MOT}
\begin{tabular}{|p{3cm}|c|c|}
\hline
&T=300K & T=300$\mu$K \\
\hline
KE & 10THz & 10MHz \\
v & 300m/s& 0.3 m/s\\
Distance traveled in 1 $\mu$s & 300$\mu$m & 0.3$\mu$m \\
\hline
\end{tabular}
\label{Tcompare}
\end{table} 

In this experiment, the $^{85}$Rb atoms are cooled by the 2nd-order ECDLs and the atoms' temperature is less than 300 $\mu$K. There are many applications for laser cooling and trapping including quantum computing \cite{Jaksch, Lukin, Gallagher3, Saffman, Tong, Singer, Liebish, Heidemann, Vogt1, Vogt2, Afrousheh, Walker1, Pierre}, simulating the condensed matter phenomena such as optical lattices \cite{Bloch}, gravity related measurements \cite{Case}, ultracold plasma \cite{Killian, Gallagher, Li}, atom lasers \cite{Ketterle2} etc.

\section{Experiment and Result}
\begin{figure}[tpb]
\centering
\includegraphics[width=3 cm,angle= 0 ,height=3 cm]{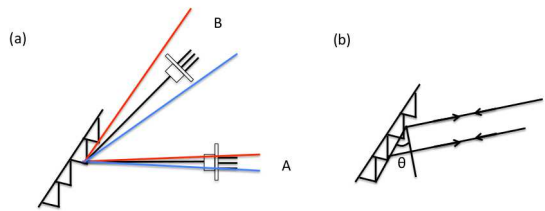}
\caption{ (a) The black beams are the direct output beams from the lasers and the feedback to the lasers. The blue and red lines correspond to the blue side and red side of the diffracted beams. If the diode is placed at position A, the first order diffracted beam from the grating is sent back to the diode. If the diode is placed at position B, the second order diffracted beam from the grating is sent back to the diode. (b) The second order diffracted beam of the input beam is directly sent back to the laser diode. The angle $\theta$ is the angle between a line parallel to the base plate of the grating and the line perpendicular to the input beam.}
\label{grating}
\end{figure}

Fig. \ref{grating} shows the schematic diagram of the external cavity laser. If the diode is placed at position A in Fig. \ref{grating}(a), the first order diffraction from the grating is sent back to the diode. If the diode is placed at position B, the second order diffracted beam from the grating is sent back to the diode. In this experiment, the laser diode is placed at position B, or we send the second order diffracted beam back to the laser diode. Since the second order is more dispersive, it is expected that the linewidth is narrower than the first order ECDLs. In this figure, the red line represents the low-frequency end of the spectra, and the blue line represents the high-frequency end of the spectra. 

We applied the 2nd-order ECDLs to our MOT setup. A commonly used ECDL is made so that the first-order diffracted beam is sent back the laser diode to form the standing-wave external cavity, and the zeroth order is the output \cite{Wieman1,Wieman2}. Very similar to those commonly used external cavity lasers, the difference between the lasers that we use and those commonly used lasers is that we send the second order diffracted beam back to the diode laser to form an external standing-wave cavity and we use the first order diffracted beam as the output. The grating used is an 830-Groove/mm gold coated grating (the part number is 43-849) purchased from Edmund optics, which is blazed at 800 nm and the blaze angle is 19$^o$23'. The laser mount is temperature stabilized. The direct output from the laser diode is about 55 mW at the current=86 mA and the temperature=14 $^o$C. If the laser polarization is vertically polarized, the power after the grating is about 32 mW, or 58$\%$ of the power is diffracted in the 1st order diffraction beam. Typically, the zero order output is used to obtain a MOT; however, we used the first order to obtain a MOT due to the high power output in the first order. It is noticed that the polarization does not significantly affect the power output. In addition, the zeroth order diffraction is about 11.5 mW, 21$\%$ of the power is diffracted in the zeroth order beam. These indicate that maximum 21$\%$ of the total power is in the second order diffraction beam, which is sent back the laser diode. The front collimation lens of the diode is an aspheric lens with 6.24 focal length, 0.4 numerical aperture, and infrared coated (A110TB-B purchased from thorlabs). We have compared the laser stability, or the frequency drifting speed, of the commonly used first-order ECDL and the second-order ECDL reported in this article. The first-order ECDL is built by switching the 830-Groove/mm grating to a commonly used 1200-Groove/mm grating with designed wavelength 500 nm (43-004 purchased from Edmund optics) \cite{Wieman1, Wieman2} and using the first-order diffraction from this 1200-Groove/mm grating. The laser drifting speeds in both configurations are not significantly different (we didn't notice the difference between those two configurations). The threshold current for the diode does not change significantly. The threshold current using the 2nd order diffraction is 24.1 mA. Unlike the previously reported results \cite{Britzger}, the threshold is 22.2 mA using the 1st-order diffraction, or the threshold current is reduced by about 2 mA by switching the 2nd order from the 830-Groove/mm grating to the 1st-order of a 1200-Groove/mm grating. If the nth order diffraction is sent back to the laser diode, from Fig. \ref{grating}(b), the angle $\theta$ can be calculated through the following equation:
\begin{equation}
\begin{split}
\theta =Sin^{-1}(\frac{n\lambda}{2d}),
\end{split}
\label{angle}
\end{equation}
where $d=\frac{1}{830 groove/mm}=1.20$ $\mu$m is the distance between the two neighboring grooves. $n$ is the number of orders. In our case, the 2nd order is sent back to the laser diode, which means $n=2$. $\lambda =780$ nm is the laser wavelength. Therefore, $\theta$=40.5$^o$ in our case. This $\theta$ is the same as the angle between the input beam and the normal of the grating.
The small change in wavelength, $\Delta \lambda$, and the small change, $\Delta\theta$, in angle are related through this equation:
\begin{equation}
\begin{split}
\Delta \lambda =\frac{2d}{n}Cos(\theta)\Delta\theta,
\end{split}
\label{angle1}
\end{equation}
or
\begin{equation}
\begin{split}
\Delta\theta=\frac{n}{2dCos(Sin^{-1}(\frac{n\lambda}{2d}))}\Delta\lambda,
\end{split}
\label{angle2}
\end{equation}
For example, if $\Delta \lambda$=1 nm, $\Delta\theta$=0.025$^o$ for the 1st-order ECDL. On the other hand,  if $\Delta \lambda$=1 nm, $\Delta\theta$=0.063$^o$ for the 2nd-order ECDL. Therefore, the second order is more dispersive. It is predicted, by using the 2nd-order ECDL, the linewidth is about half of the linewidth of the 1st-order ECDL or narrower.

\begin{figure}[tpb]
\centering
\includegraphics[width=3 cm,angle= 0 ,height=3 cm]{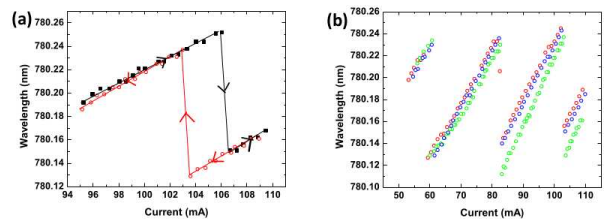}
\caption{ (a) Wavelength as a function of diode current. The black squares are measured by increasing the current, and the red circles are measured by decreasing current. The lines and arrows show how the current is changed and is for viewing purposes. (b) Three sets of data measured by decreasing the current. The uncertainty of the current is 0.01 mA, and the uncertainty of the wavelength is about 3 pm.}
\label{diode}
\end{figure}

We characterized these 2nd-order ECDLs that we used. As we change the current, the laser output wavelength changes. Typically, if we increase the current, the laser wavelength increases. The diode laser's mode hops when it reaches a certain wavelength. In other words, as we keep increasing the current, the wavelength drops to a lower value, this is called mode hopping. After the mode hopping, the wavelength increases again as we increase the current as shown in Fig. \ref{diode}. We also observed the hysteresis effect, similar to the hysteresis effect observed in magnetism, in diode lasers. The wavelength is measured by a wavemeter purchased from MOG lab. From the wavelength vs. current plot in \ref{diode}(a), it is shown that there is a hysteresis effect. This effect is more obvious when the current is high. The red circles are measured right after the black squares. In other words, we first increase the current as shown by the black squares and then decrease the current as shown by the red circles (the arrows also indicate the current changing direction). The fact that this effect is reduced after waiting a long time indicates that this effect is caused by the thermalization of the diode. From Fig. \ref{diode}(b), it is shown that the data points are more separated at higher currents, which indicates that the current fluctuation is higher at a higher current. 
More importantly, it is noticed that the wavelength changes in steps before the big mod hops at 103.5 mA and 106.5 mA shown in Fig. \ref{diode}. The step size is about 6 pm. As shown below, this is consistent with the free spectrum range of the external cavity. The free spectrum range of a cavity, $\Delta f$, is defined as
\begin{equation}
\begin{split}
\Delta f=\frac{c}{nL},
\end{split}
\label{free_spectrum_range}
\end{equation}
where $f$ is frequency, $c$ is the speed of light, $n$ is the index of refraction of the medium within the cavity, and $L$ is the round-trip cavity length (twice of the physical length). If we apply, $\lambda f=c$, where $\lambda$ is wavelength, the following equation is obtained.
\begin{equation}
\begin{split}
\Delta \lambda=\frac{\lambda^2}{nL},
\end{split}
\label{free_spectrum_range_2}
\end{equation}
The external round-trip cavity length, $L$, in this experiment is about 10 cm. If we set the index refraction of air $n=1$, the $\Delta \lambda$ calculated from Eq. \eqref{free_spectrum_range_2} is 6.1 pm, which is quite consistent with the measured step size, 6 pm. This means that the step size, which is critical for the saturated absorption spectrum (if a smaller step size is more beneficial),
 can be reduced by increasing the external cavity length. In other words, the small wavelength gaps, 6 pm, can be reduced, so that no angle adjustment is needed, or the wavelength can be purely tuned by the current. The small wavelength gap, 6 pm, is not a problem in our experiments since the external grating's angle can be adjusted to produce a continuous output. Here the small step size, 6pm, is different from mode-hopping. The mode hopping size is about 120 pm as discussed below. The big hop is caused by the confinement of the internal cavity of the diode. From the hopping wavelength range, the hopping happens at 103.5 mA and 106.5 mA in Fig. \ref{diode}(b), the thickness of the laser diode can be estimated. The typical index of refraction for semiconductors is about 3.5. From Fig. \ref{diode}(b), the $\Delta \lambda$, or continuous range, is about 0.12 nm, or 120 pm, which indicates that the thickness, or the internal round-trip cavity length $L$, of the laser diode is about 1.5 mm calculated from Eq. \eqref{free_spectrum_range_2}. In other words, the physical length of the laser diode, or the thickness of the semiconductor of the diode, is about 0.75 mm. This shows that to reduce the laser hopping issues raised here or increase the tunable range, the thickness of the laser diode needs to be reduced. This information may be useful for the laser diode manufactures to make more user-friendly diodes for laser cooling and trapping.

\begin{figure}[tpb]
\centering
\includegraphics[width=6 cm,angle= 0 ,height=6 cm]{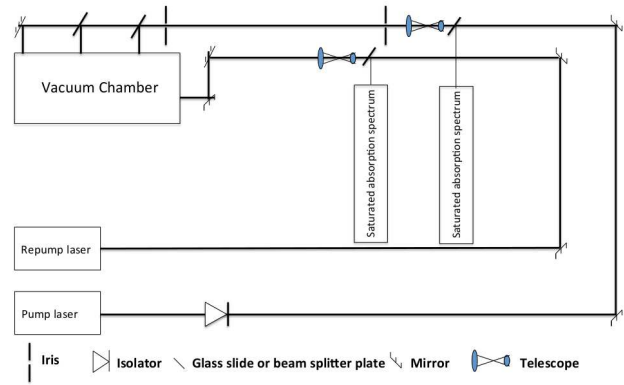}
\caption{ The experimental setup for the MOT setup. The names of different components are shown on the bottom of the figure. }
\label{setup}
\end{figure}

The MOT setup is very similar to the common setup widely used in the science community. Here is a brief description of the MOT setup that we built. Fig. \ref{setup} shows the optical diagram for this experiment. 
For example, we did not use an isolator in the repump laser. Instead, we made the optical path a lot longer than the optical path in other MOT setups to reduce the amount of reflection from mirrors and lenses downstream. In addition, we expanded the repump beam to make sure it covers the MOT area.  The maximum diameter of the trapping laser is about 8 mm, and the maximum diameter of the repump laser is about 10 mm. The sizes are measured by looking at the profile of the laser beam instead of the knife edge method. Due to the non-uniform distribution of the laser power, or the dark spot, in the repump laser, we did not use the knife edge method. One potential problem with a long optical path is that the slight change in the direction of the laser output will mess up the alignment. We could use two pinholes in the repump beam like what we did in the trapping beam to help with the alignment. However, it turns out that adding two pinholes is not necessary for our experiment because of the large size of the repump beam. The output power from the trapping laser and repump laser is about 30 mW. The output from the trapping laser is reflected from two infrared-coated mirrors. About 15$\%$ of the beam is then reflected from the front and back surface of a glass slide and sent through the saturated absorption spectrum setup for the trapping laser. The beam that passes through the glass slide will be magnified by a telescope, and then passes through two pinholes or two irises. The beam is then equally split into three beams, about 33$\%$ in each beam, and sent to the vacuum chamber to do the three-dimensional cooling (3D cooling). In addition to the trapping beams, a repump beam created from the repump laser is required to reduce the loss of the trapped atoms. 

\begin{figure}[tpb]
\centering
\includegraphics[width=3 cm,angle= 0 ,height=3 cm]{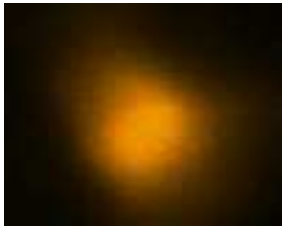}
\caption{ The image of the magneto-optical trap (MOT) created by the 2nd-order ECDLs.}
\label{MOT}
\end{figure}

Fig. \ref{MOT} shows the image of the MOT. This image is taken by a Stingray F-145C camera and the software that we used for taking the image is Vimba Viewer. The total number of atoms is about 10$^6$. In some cases, the MOT shape is not the typical Gaussian distribution, which may be caused by the shape of the overlapping area of the lasers. In addition, different shapes can be used to study quantum phenomena, such as quantum Hall effects, and quantum dots. Getting different shapes indicates that it is possible to shape the MOT to carry out those experiments. 

\section{Conclusion}
In summary, we have built the 2nd-order external-cavity lasers, in which we sent the second order diffracted laser beam back into the laser diodes. Moreover, we have observed that the wavelength of the diode lasers changes in steps as the current increases. This non-continuous dependence is caused by the standing wave cavity. It is shown that the wavelength vs. laser current measurements can be used to estimate the size of the thickness of the gain medium or the thickness of internal cavity of the laser diode. To reduce the laser hopping issues raised here or increase the tunable range, the thickness of the laser diode needs to be reduced. The wavelength vs. laser current measurements can be used to accurately measure the cavity size in general. This information may be useful for the laser diode manufactures to make more user-friendly diodes for laser cooling and trapping.
In addition, the external cavity diode lasers show the hysteresis effect caused by thermal effects. Furthermore, the MOT achieved using the 2nd-order diode lasers was characterized. 

\section{Acknowledgement}

It is a pleasure to acknowledge valuable discussions with Drs. Duncan Tate,  Kevin Wright, and Seth Aubin. This work was supported by Army Research Office, the University of South Alabama faculty development council (USAFDC), DOE/EPSCOR, and the University of South Alabama.

\end{document}